\documentclass[conference]{IEEEtran}
\usepackage{cite}
\usepackage[hidelinks]{hyperref}
\usepackage{amsmath,amssymb,amsfonts}
\usepackage{graphicx}
\usepackage{textcomp}
\usepackage{xcolor}
\usepackage{url}
\usepackage{booktabs}
\usepackage{multirow}
\usepackage{balance}
\usepackage{array}
\usepackage{colortbl}
\usepackage[most]{tcolorbox}
\def\BibTeX{{\rm B\kern-.05em{\sc i\kern-.025em b}\kern-.08em
    T\kern-.1667em\lower.7ex\hbox{E}\kern-.125emX}}
    \makeatletter
\newcommand{\linebreakand}{%
  \end{@IEEEauthorhalign}
  \hfill\mbox{}\par
  \mbox{}\hfill\begin{@IEEEauthorhalign}
}
\makeatother

\usepackage{makecell}

\definecolor{hdrblue}{RGB}{31,56,100}
\definecolor{rowA}{RGB}{220,230,241}
\definecolor{rowB}{RGB}{235,243,251}

\def\BibTeX{{\rm B\kern-.05em{\sc i\kern-.025em b}\kern-.08em
    T\kern-.1667em\lower.7ex\hbox{E}\kern-.125emX}}

\begin{document}

%─────────────────────────────────────────────────────────────────────────────
\title{Lifecycle-Integrated Security for AI-Cloud Convergence
in Cyber-Physical Infrastructure}

\author{
\IEEEauthorblockN{S M Zia Ur Rashid}
\IEEEauthorblockA{\textit{Dept. of Electrical and Computer Engineering}\\
The University of Tulsa\\ Tulsa, OK, USA\\
Email: smziaurrashid@gmail.com}
\and
\IEEEauthorblockN{Deepa Gurung}
\IEEEauthorblockA{\textit{Dept. of Business Administration}\\
Joongbu University\\ Seoul, South Korea\\
Email: grgdips110@gmail.com}
\and
\IEEEauthorblockN{Sonam Raj Gupta}
\IEEEauthorblockA{\textit{Independent Researcher}\\
Tulsa, OK, USA\\
Email: guptasonamraj97@gmail.com}

% This creates the second row
\and
\IEEEauthorblockN{\\} % This acts as a spacer for the left column
\linebreakand
\IEEEauthorblockN{Suman Rath}
\IEEEauthorblockA{\textit{Dept. of Electrical and Computer Engineering}\\
The University of Tulsa\\ Tulsa, OK, USA\\
Email: suman-rath@utulsa.edu}
}

% \author{
% \IEEEauthorblockN{S M Zia Ur Rashid}
% \IEEEauthorblockA{\textit{Dept.\ of Electrical and Computer Engineering}\\
% The University of Tulsa, Tulsa, OK, USA\\
% Email: smziaurrashid@gmail.com}
% \and
% \IEEEauthorblockN{Deepa Gurung}
% \IEEEauthorblockA{\textit{Dept.\ of Business Administration}\\
% Joongbu University, Seoul, South Korea\\
% Email: grgdips110@gmail.com}
% \and
% \IEEEauthorblockN{Sonam Raj Gupta}
% \IEEEauthorblockA{\textit{Independent Researcher}\\
% Tulsa, OK, USA\\
% Email: guptasonamraj97@gmail.com}
% \and
% \IEEEauthorblockN{Suman Rath}
% \IEEEauthorblockA{\textit{Dept.\ of Electrical and Computer Engineering}\\
% The University of Tulsa, Tulsa, OK, USA\\
% Email: suman-rath@utulsa.edu}
% }
% Source - https://tex.stackexchange.com/a/618225
% Posted by Engr. Nnoli K. Pal, modified by community. See post 'Timeline' for change history
% Retrieved 2026-02-24, License - CC BY-SA 4.0

% \IEEEoverridecommandlockouts
% \IEEEpubid{\makebox[\columnwidth]{979-8-3315-7310-2/26/\$31.00~\copyright2026 IEEE \hfill}
% \hspace{\columnsep}\makebox[\columnwidth]{ }}

\maketitle

\IEEEpubidadjcol

%─────────────────────────────────────────────────────────────────────────────
\begin{abstract}
The convergence of Artificial Intelligence (AI) inference pipelines with cloud infrastructure
creates a dual attack surface where cloud security standards and AI
governance frameworks intersect without unified enforcement mechanisms. AI governance, cloud security, and
industrial control system standards intersect without unified enforcement, leaving hybrid deployments exposed to cross-layer attacks that threaten safety-critical operations. This paper makes three primary contributions: (i) we synthesize these frameworks into a lifecycle-staged threat taxonomy structured around explicit attacker capability tiers, (ii) we
propose a Unified Reference Architecture spanning a Secure Data Factory, a hardened model supply chain, and a runtime governance layer, (iii) we present a case study through Grid-Guard, a hybrid Transmission System
Operator scenario in which coordinated defenses drawn from NIST AI~RMF,
MITRE ATLAS, OWASP AI Exchange and GenAI, CSA MAESTRO, and NERC CIP defeat a multi-tier
physical-financial manipulation campaign without human intervention.
Controls are mapped against all five frameworks and current NERC CIP
standards to demonstrate that a single cloud-native architecture can simultaneously satisfy AI governance,
adversarial robustness, agentic safety, and industrial regulatory
compliance obligations.
% The convergence of Artificial Intelligence (AI) inference pipelines with hybrid cloud infrastructure in safety-critical cyber-physical systems creates a governance gap: existing AI security, cloud security, and industrial control standards address overlapping risk surfaces but lack unified lifecycle enforcement. This fragmentation leaves hybrid OT–cloud deployments exposed to data poisoning, supply chain compromise, latency-based resource exhaustion, and unsafe agentic actions. We propose a lifecycle-integrated Unified Reference Architecture that combines cryptographic provenance, physics-aware validation, adversarially hardened development pipelines, and runtime policy-enforced governance. The architecture is analytically evaluated through a scenario-driven Transmission System Operator case study (“Grid-Guard”) that models coordinated multi-tier adversarial campaigns. Validation criteria include deterministic threat containment, bounded runtime latency, fail-safe degradation to on-premise control, and concurrent cross-framework compliance mapping. The analysis demonstrates that a single cloud-native architecture can reconcile AI governance, adversarial robustness, and NERC CIP reliability obligations without bespoke per-framework engineering.
\end{abstract}


%─────────────────────────────────────────────────────────────────────────────
\section{Introduction}

Critical infrastructure sectors such as power grids, manufacturing, and
logistics are migrating from isolated on-premise deployments toward
hybrid and cloud-native architectures to support AI-driven telemetry
analytics, predictive maintenance, and automated real-time control~\cite{b2,b3}. 
IEEE PES Task Force documents this
shift comprehensively, identifying elastic compute, large-scale Advanced Metering Infrastructure (AMI) or Phasor Measurement Unit (PMU)
data management, and data-driven AI modeling as primary business
drivers---while simultaneously identifying cybersecurity and regulatory
compliance as the leading barriers to adoption \cite{b1}. What
makes this difficult is that cloud, AI, and industrial OT security
developed independently under incompatible assumptions, and integrating
them without a coherent security architecture creates vulnerability gaps
that none of the three disciplines individually anticipates\cite{b5}. The
problem is most visible in real-world when multi-tenant cloud platforms
process simultaneous telemetry from AMI infrastructures, industrial IoT
sensors, and distributed monitoring networks at scale.

Although adversarial ML defenses\cite{b7,b8,b9,b11}, Zero Trust architectures\cite{b13}, and NERC CIP \cite{b12} compliance standards have each matured independently, they remain operationally fragmented in hybrid OT–cloud deployments\cite{b1}. Existing frameworks address portions of the AI lifecycle or infrastructure stack, but none integrates lifecycle AI security, cloud-native enforcement, and physics-aware validation within a unified architecture suitable for safety-critical systems. This fragmentation creates enforcement gaps, overlapping compliance burdens, and unresolved timing constraints in deterministic control environments.
This paper addresses these gaps through the following contributions:
\begin{itemize}
  \item We develop a lifecycle-staged threat taxonomy organized
  around three attacker capability tiers, synthesizing MITRE ATLAS,
  OWASP, CSA MAESTRO, and NIST AI~RMF into a unified threat landscape
  for AI-Cloud CPS deployments.
  % \item We introduce a \textbf{physics-aware data pipeline} that couples
  % cryptographic SPIFFE provenance with power system swing-equation
  % consistency checks, providing a detection layer that purely statistical
  % defenses cannot replicate.
  \item We propose a Unified Reference Architecture spanning a
  Secure Data Factory, a hardened model supply chain, and a Governance
  Sidecar with a Latency Circuit Breaker for
  runtime agentic safety.
  \item We present Grid-Guard, a scenario-based case study demonstrating layered defense execution and concurrent cross-framework compliance.
  % \item We present Grid-Guard, a scenario-based design study
  % that traces representative attack scenarios through each defensive
  % layer and illustrates concurrent cross-framework compliance without
  % bespoke engineering for any individual standard.
\end{itemize}
The remainder of the paper is structured as follows. Section II surveys related work. Section III presents the unified threat taxonomy. Sections IV–VI detail the three framework phases. Section VII presents the case study and compliance mapping. Section VIII concludes.
\section{Background and Related Work}

\subsection{Cloud and Hybrid Adoption}
The IEEE PES Task Force identifies elastic compute, scalable AMI/PMU storage, and ML/AI tooling as primary drivers for utility cloud adoption, documenting the hybrid model—cloud for variable workloads, on-premise for safety-critical control—as the dominant pattern ~\cite{b1}. However, NERC CIP-005-8 and CIP-007-7.1 presuppose static perimeters that cloud auto-scaling and dynamic migration violate, creating the compliance gap this work addresses ~\cite{b12}.

% The IEEE PES Task Force on Cloud Computing in Power Systems identifies
% three operational imperatives driving cloud adoption in utilities:
% elastic compute for large-scale grid simulation, scalable storage for
% AMI and PMU data streams, and accessible ML/AI tooling for data-driven
% forecasting and anomaly detection. The Task Force
% documents (PES TR-92) the hybrid cloud model---cloud for variable, burst workloads;
% on-premise for fixed, safety-critical control---as the dominant
% deployment pattern for utilities navigating regulatory constraints \cite{Zhang_TSG2022}. However, NERC CIP-005-8 and CIP-007-7.1 presuppose
% static, enumerable perimeters; cloud auto-scaling and dynamic workload
% migration violate both assumptions simultaneously, producing the
% compliance gap that this work addresses \cite{b7,b8}.

In hybrid deployments, deterministic on-premise controllers (e.g., SCADA and protection relays) operate separately from cloud-based AI analytics, creating timing asymmetries. Probabilistic AI outputs must integrate with control loops that require strict, deterministic timing guarantees. 
Conventional mitigation strategies including sandboxing and rate-limiting do not applicable directly to CPS environments where timing guarantees are safety-critical rather than performance-oriented ~\cite{b3,b4}.

\subsection{Adversarial Machine Learning (AML)}
% AI models deployed within cloud-hosted infrastructure analytics are vulnerable to AML attacks. 
MITRE ATLAS classifies AML threats including data poisoning, model extraction, and supply chain compromise \cite{b10}. In cyber-physical systems (CPS) settings, adversaries exploit the
continuous, high-volume telemetry ingestion documented in PES TR-92 to
inject physically infeasible measurements that evade statistical
detectors \cite{b1}. Current AML defenses mostly on statistical anomaly detection and adversarial training, which rarely incorporate physical process constraints, leaving critical infrastructure analytics vulnerable to physics-unaware data manipulation \cite{b9}.

\subsection{Agentic AI and Hybrid Operational Architectures}
As AI systems gain autonomy and begin executing actions rather than just making predictions, we encounter a new class of operational risk. OWASP GenAI (LLM Top 10) \cite{b8} and CSA MAESTRO \cite{b11} frameworks discuss risks in application layers and agentic ecosystems, yet neither provides actionable guidance on building trust boundaries using cloud-native security primitives in industrial contexts.

% \subsection{Integrated AI–Cloud Security Frameworks}
Recent integrated framework work has
improved intrusion detection and model governance in cloud contexts
\cite{b11,b7,b8}, but three structural gaps persist: (1)~no framework
simultaneously addresses AI lifecycle security, cloud infrastructure
protection, and OT physical safety; (2)~physics-aware validation is
absent from all AI security frameworks; and (3)~no framework provides
operational guidance for hybrid OT/cloud deployments. These gaps
motivate the unified architecture developed in this work.

\section{Unified AI-Cloud Threat Landscape}

\begin{figure*}[t]
\centering

\begin{tcolorbox}[
colback=blue!4!white,
colframe=blue!40!black,
arc=3mm,
boxrule=0.4pt,
width=0.90\textwidth
]

\centering
\includegraphics[width=\textwidth]{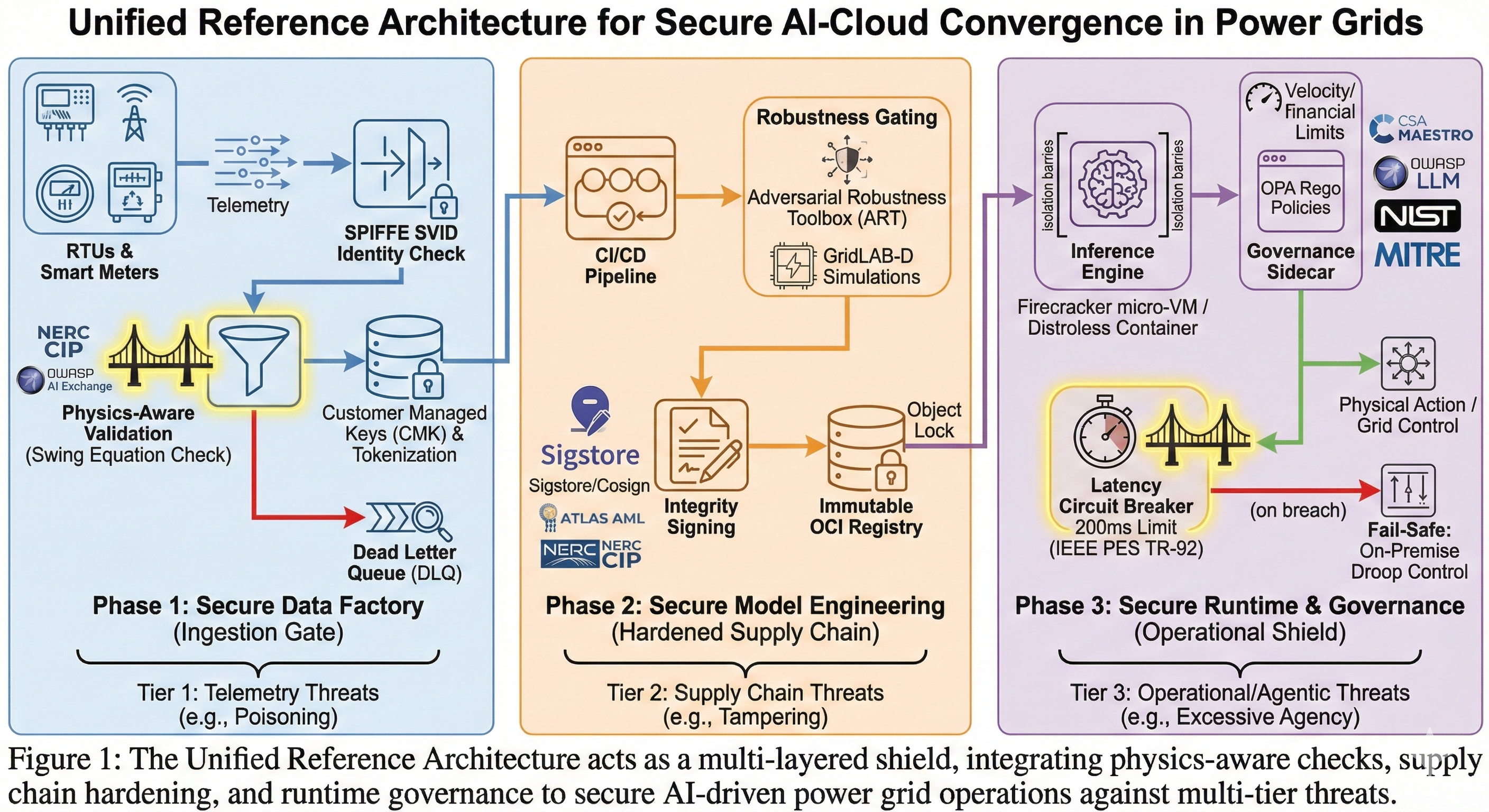}

\end{tcolorbox}

\caption{\textbf{Unified Reference Architecture}}
\label{fig:framework}
\end{figure*}

We structure threats across three lifecycle-aligned vectors, each
associated with a distinct attacker capability tier as depict in Fig.~\ref{fig:framework}. Every control
introduced in Sections IV--VI is designed against the specific access
requirement of the relevant tier.

\textbf{Attacker Capability Tiers.}
\textit{Tier~1} adversaries hold persistent access to a telemetry
source---a compromised substation RTU, a malicious insider at a meter
aggregation point, or a spoofed edge device---enabling sustained,
low-and-slow injection of malicious data without triggering access
alerts.
\textit{Tier~2} adversaries control a stage in the model development
or distribution pipeline---CI/CD credentials, a package registry
account, or a dependency namespace---enabling artifact substitution or
poisoning before deployment.
\textit{Tier~3} adversaries operate exclusively through the inference
API---submitting crafted queries, injecting adversarial prompts, or
manipulating agentic tool-use workflows---without any privileged access
to the underlying infrastructure.

\subsection{Data-Centric Threats (Tier 1)}
Tier~1 adversary exploits the AMI/PMU telemetry ingestion surface
\cite{b1}. 
\textbf{Data and model
poisoning} injects inputs statistically
indistinguishable from legitimate telemetry but violating physical
laws---by construction, this attack is designed to pass the statistical
anomaly detectors on which most AI security defenses rely. \textbf{RAG
and context store injection} targets retrieval databases to skew agent reasoning at query time,
leaving the underlying model untouched and the attack harder to detect. \textbf{Sensitive
information disclosure} through membership inference on model outputs can also expose BCSI without
any direct access to training data.

% \textbf{Data and model
% poisoning} injects inputs statistically
% indistinguishable from legitimate telemetry but violating physical
% laws---by construction, this attack is designed to pass the statistical
% anomaly detectors on which most AI security defenses rely. \textbf{RAG
% and context store injection} compromises vector databases used for
% retrieval-augmented generation, manipulating agent reasoning at
% retrieval time without modifying the model itself. \textbf{Sensitive
% information disclosure} extracts BCSI
% from model outputs through membership inference and output analysis.

\subsection{Model-Centric Threats (Tier 2)}
Tier~2 adversaries work within the model development and distribution
pipeline. \textbf{Supply chain compromise} may replace verified artifacts or models with backdoored alternatives. When it comes to grid operations, a compromised forecasting model can deliberately underestimate the demand, thus causing the load imbalances.
\textbf{Improper output handling} allows unsafe model-generated recommendations such as physically infeasible dispatch instructions to reach downstream control mechanisms. \textbf{Model extraction}
enables downstream white-box adversarial exploitation of proprietary
operational models.

% Tier~2 adversary targets the model development and distribution
% pipeline. \textbf{Supply chain compromise} substitutes validated artifacts with
% trojaned versions---in grid operations, a compromised forecasting model
% could systematically under-report demand to induce load imbalance.
% \textbf{Improper output handling}
% occurs when unsafe model outputs (e.g., physically infeasible dispatch
% recommendations) propagate to downstream control systems without
% validation. \textbf{Model extraction}
% enables downstream white-box adversarial exploitation of proprietary
% operational models.

\subsection{Operational and Agentic Threats (Tier 3)}
\textbf{Unbounded Consumption} induces inference resource exhaustion, producing both
financial Denial of Wallet in cloud billing and, critically, latency
injection that violates AGC sub-second timing requirements
\cite{b1,b11}. \textbf{Excessive Agency} enables agentic systems
manipulated through prompt injection or goal misalignment to execute
unsafe load shedding or unconstrained market orders through legitimate
channels. \textbf{System prompt leakage} exposes the operational context and safety constraints embedded in
agent system prompts, enabling targeted adversarial bypasses.
\textbf{Container escape and RCE}
exploit inference library vulnerabilities for lateral movement into
the cloud environment.

%─────────────────────────────────────────────────────────────────────────────
\section{Phase 1: Secure Data Factory}

\subsection{Cryptographic Provenance and Ingestion Gating}
% All telemetry is treated as untrusted by default (Zero Trust; NIST
% AI~RMF Govern 1.2; OWASP AI Exchange general control: supply chain
% management). SPIFFE assigns cryptographic SVID identities to all data producers (RTUs, smart meters, edge devices). 
% All telemetry is treated as untrusted by default, aligning with Zero Trust principles (NIST AI RMF Govern 1.2; OWASP AI Exchange supply chain management controls). RTUs, smart meters, and edge nodes are all possible points of failure, and any ingestion design
% that relies on network location for trust will fail against an insider
% or a spoofed device. To mitigate this, we adopt SPIFFE-based workload
% identity, which assigns each data producer a short-lived cryptographic
% credential (SVID) that the ingestion gateway can verify independently
% of network topology. 
Field devices including RTUs, smart meters, and edge nodes
are all potential compromise points, so the ingestion gateway
cannot rely on network location as a trust signal. We use
SPIFFE workload identity to assign each data producer a short-lived cryptographic credential (SVID). Telemetry that arrives without a valid SVID is sent to a Dead Letter Queue (DLQ) for forensic retention to meet the security event logging requirements as defined in CIP-007-7.1 R~5 and the monitoring objectives defined in CIP-015-1.

Tenant isolation is enforced through per-tenant Customer Managed Keys (CMK), ensuring that compromise in one tenant's pipeline cannot expose BES Cyber System Information of others, addressing CIP-011-4.1~R1 and mitigating the risk identified in LLM02:2025.
% Tenant isolation is enforced through per-tenant Customer Managed Keys (CMK)
% rather than logical namespace separation alone. CMK based separation ensures that compromise or misconfiguration in one tenant's pipeline does not expose the BES Cyber System Information of other tenants, thus directly addressing the provisions of CIP-011-4.1~R1, mitigating the risk identified in LLM02:2025.

% Together, the SVID verification layer
% and CMK isolation implement a Zero Trust ingestion boundary consistent
% with NIST AI~RMF Govern~1.2, where trust is earned through
% cryptographic attestation rather than assumed from network position.

% SPIFFE provides cryptographic SVID identities for RTUs, smart meters, and edge devices. Cloud ingestion
% gateway operates as a Zero Trust Policy Enforcement Point: telemetry
% lacking valid SPIFFE signatures is routed to a Dead Letter Queue (DLQ) for forensic analysis, preserving CIP-007-7.1 R5 and CIP-015-1 audit requirements. Customer Managed Keys (CMK) per tenant provide logical isolation, addressing OWASP LLM02:2025 and CIP-011-4.1~R1 requirements through data separation.

\subsection{Physics-Aware Validation and Privacy Preservation}
% Authenticated telemetry passes through two-stage sanitization addressing OWASP LLM04:2025 (Data and Model Poisoning) and MAESTRO L2 (Data Operations) threats. 
Authenticated telemetry undergoes a two-stage sanitization pipeline
that addresses data and model poisoning, and data
operations threats. \textit{Stage~1} validates schema compliance and monitors KL divergence against baselines.
% \textit{Stage~1} applies
% statistical validators enforcing schema compliance and KL divergence
% monitoring against operational baselines. 
\textit{Stage~2} implements Physics Consistency Checks by cross-referencing frequency measurements against the swing equation,
\begin{equation}
  \Delta f \approx -\frac{\Delta P}{2H \cdot f_0},
  \label{eq:swing}
\end{equation}
where $H$ is system inertia and $f_0$ is nominal frequency. Any batch which has a reported deviation not compatible with PMU-measured inertia at reference nodes is rejected regardless of passing statistical filters. This requirement can be used to detect physics-mimicry attacks designed to look realistic in Stage 1 (LLM04:2025; MAESTRO L2). Format-Preserving Encryption (FPE) is used to tokenise sensitive data fields to meet the requirements of CIP-011-4.1~R1.2 and still retain the analytical utility of the datasets.

% Telemetry batches which have deviations not consistent with PMU-measured inertia at reference nodes are rejected regardless of their statistical
% plausibility. This provides a detection surface that is structurally
% inaccessible to Tier~1 physics-mimicry attacks, whose defining
% characteristic is statistical legitimacy. Sensitive fields are
% tokenized via Format-Preserving Encryption (FPE), satisfying
% CIP-011-4.1~R1.2 while preserving the analytical utility of the data.

% Sensitive fields are tokenized using Format-Preserving Encryption (FPE), satisfying
% CIP-011-4.1~R1.2 and OWASP LLM02:2025 disclosure controls while preserving analytical utility.

%─────────────────────────────────────────────────────────────────────────────
\section{Phase 2: Secure Development \& Supply Chain}

\subsection{Adversarial Robustness in the CI Pipeline}
The model robustness is imposed as a mandatory continuous-integration gate further mitigating the LLM04:2025 and MAESTRO L1/L3 threats. Training data is augmented with Fast Gradient Sign Method (FGSM) and Projected Gradient Descent (PGD) perturbations. For CPS deployments, physics-based simulators, e.g. GridLAB-D and PSCAD are used to generate physically realizable fault situations as described by the safe-operations guide within PES TR-92 ~\cite{b1}. Each candidate is
benchmarked against the production baseline using the Adversarial
Robustness Toolbox (ART) ~\cite{b6}; candidates achieving higher
accuracy but lower robustness fail automatically. Both metrics are
logged as provenance metadata (NIST AI~RMF Measure 2.5/2.6;
CIP-010-5~R1; MAESTRO L5: Evaluation and
Observability).

\subsection{Model Artifact Integrity}
Model artifacts are signed immediately upon validation using
Sigstore/Cosign with TPM-backed keys, covering the binary, metadata
manifest, and timestamp---directly countering LLM03:2025 and MAESTRO L4 threats. Signed
artifacts are stored in an OCI-compliant immutable registry with Object
Lock (write-once) semantics, eliminating the TOCTOU window between
validation and deployment. Provenance metadata (signing identity,
timestamp, CI run~ID) satisfies CIP-010-5~R1.1/R3 and CIP-013-3~R1.2.5
supply chain authenticity requirements; NIST AI~RMF Govern~3.1 is
satisfied through documented artifact accountability chains.

%─────────────────────────────────────────────────────────────────────────────
\section{Phase 3: Secure Runtime \& Governance}

\subsection{Runtime Isolation  and Canary Deployment}
% Inference services execute in distroless containers or Firecracker
% micro-VMs containing only the application binary---no shells, package
% managers, or OS utilities. An adversary achieving RCE via an inference
% library vulnerability finds no execution surface for lateral movement
% (MAESTRO L4; OWASP LLM05:2025). 

Inference services execute in distroless containers or Firecracker micro-VMs with only the application binary, eliminating execution surfaces for lateral movement following RCE exploits (MAESTRO L4; LLM05:2025). A service mesh (Istio/Linkerd) enforces mTLS with SPIFFE workload
identities; ingress is restricted to the API Gateway and Governance
Sidecar; egress is whitelisted exclusively to designated on-premise
SCADA connectors via VPN/Direct Connect. This implements the CIP-005-8
R1 EAP requirement at the cloud-to-OT trust boundary, with the mesh
audit log satisfying CIP-007-7.1~R5 and CIP-015-1 (Internal Network
Security Monitoring) for anomaly and lateral movement detection.

Model updates are promoted through a canary strategy covering 5 to
10 percent of live telemetry, tracking semantic drift via KL divergence
and physical constraint violations where frequency predictions fall
outside the 59 to 61~Hz band or dispatch exceeds rated capacity. Either of the conditions will trigger an automatic rollback before going to production, and satisfying the fulfillment of the requirements, defined in CIP-010-5 R1.1 and NIST AI RMF Manage 2.2.

% Either condition triggers automatic rollback before promotion to
% production, satisfying CIP-010-5~R1.1 and NIST AI~RMF Manage~2.2
% (LLM09:2025; MAESTRO L5).

% Either condition triggers automatic rollback (LLM09:2025; MAESTRO L5;
% CIP-010-5~R1.1; NIST AI~RMF Manage~2.2).

% \subsection{Physics-Aware Canary Deployment}
% Model updates are promoted through a canary strategy (5--10\% of live
% telemetry), addressing OWASP LLM05:2025 (Improper Output Handling)
% and MAESTRO L5 (Evaluation and Observability) simultaneously. The
% monitoring service tracks two CPS-specific failure modes:
% (1)~\textit{semantic drift} via KL divergence between candidate and
% production output distributions, and (2)~\textit{physical constraint
% violations} (e.g., frequency predictions outside the 59--61\,Hz
% operational band, or dispatch exceeding rated asset capacity). A
% candidate producing statistically normal but physically infeasible
% outputs---OWASP LLM09:2025 Misinformation in a safety-critical
% context---is more dangerous than one producing obvious anomalies, since
% it would pass conventional monitoring. Both conditions trigger automatic
% rollback, satisfying CIP-010-5~R1.1 pre-deployment testing requirements
% and NIST AI~RMF Manage~2.2.

\subsection{Governance Sidecar and Latency Circuit Breaker}

Manipulating the logic of an authorized agent allows them to cause operational harm while remaining within their nominal permission scope, such as when they place market bids or send dispatch directives. The Governance Sidecar inspects all outbound tool calls and compares them to OPA Rego policies before executing them. It operates under a dedicated SPIFFE workload
identity and loads its policy bundle solely from the immutable registry,
so a compromised orchestrator cannot alter enforcement at runtime
(MAESTRO L3/L6; AML.T0051; LLM06:2025). Sidecar failure defaults to
deny-all. The imposed limits include a velocity limit of five signals per minute, a financial limit of fifteen percent in relation to the one-hour rolling average, and a scope of least-privilege actions as per NIST SP 800-207 ~\cite{b13}. All denied actions are logged to satisfy the audit trail requirements of CIP-007-7.1~R5 and CIP-015-1.

Furthermore, the architecture incorporates a Latency Circuit Breaker to mitigate sponge attacks by preventing inflated inference delays from violating the sub-second response requirements of Automatic Generation Control (AGC). If the threshold is exceeded, the AI agent is automatically decoupled from the control loop before grid stability is affected. Its 200~ms budget
is derived from PES TR-92 AGC requirements~\cite{b1}
by partitioning the 400~ms AGC loop budget equally between AI inference
and network transit plus actuator response. When the threshold is
exceeded, HTTP~503 is issued and frequency regulation transfers
immediately to on-premise droop control. A latency-stalled model and a
failed sensor are treated as operationally equivalent (LLM10:2025;
AML.T0029).
\section{Grid-Guard: Reference Architecture}
Grid-Guard applies the Unified Reference Architecture to a
hybrid Transmission System Operator (TSO) setting. Coordinated multi-tier attack phases are injected conceptually, and defensive responses are traced across architectural layers. The evaluation does not rely on field deployment but on structured threat modeling, control trace analysis, and cross-framework compliance mapping as presented in Table~\ref{tab:compliance_matrix}.

% It walks through representative
% attack scenarios to show how each defensive layer responds and where
% the controls satisfy concurrent framework requirements. 
% Table~\ref{tab:compliance_matrix} presents
% the full control-to-framework alignment.
% The preceding sections defined the Unified Reference Architecture in
% the abstract. Grid-Guard instantiates and stress-tests it through a
% hybrid Transmission System Operator (TSO) deployment that serves as
% both a concrete implementation guide and a cross-framework validation
% vehicle. The central claim is that a single, coherent set of cloud-native
% security controls can simultaneously satisfy NIST AI~RMF, MITRE ATLAS,
% OWASP AI Exchange, OWASP LLM Top~10:2025, CSA MAESTRO, and NERC CIP
% concurrently---without bespoke compliance engineering for each framework.
% Table~\ref{tab:compliance_matrix} captures this alignment; the
% subsections below provide the operational narrative that grounds it
% in physical and regulatory reality.

\subsection{Deployment Architecture and Compliance Scoping}
The TSO runs load forecasting models, AMI analytics pipelines, and an
agentic dispatch optimizer on cloud infrastructure. NERC CIP-002-5.1a describes the above workloads as medium-impact Bulk Electric System (BES) Cyber Systems with an External Routable Connectivity, which means that they will be subject to full compliance framework within CIP-005-8 to CIP-015-1. Protection relays, RTUs, and AGC remain on-premise within NERC CIP ESP boundaries. Therefore, the interface between cloud services and on-premises infrastructure is the major point of enforcing the trust in this operational model.

\subsection{Threat Scenario: The Blackout-Bankrupt Attack}
%─────────────────────────────────────────────────────────────────────────────
% COMPLIANCE TABLE
%─────────────────────────────────────────────────────────────────────────────
\begin{table*}[!ht]
\centering
\caption{Unified Compliance Matrix: Threat Vectors, Controls, and
Cross-Framework Alignment.} 
% OWASP references use 2025 LLM Top~10
% numbering \cite{b46}. NERC CIP versions per FERC-725B, Feb.\ 2026
% \cite{FERC725B_2026}. MAESTRO layers: L1=Foundation Models,
% L2=Data Ops, L3=Agent Frameworks, L4=Deploy \& Infra,
% L5=Eval \& Observability, L6=Security \& Compliance, L7=Agent Ecosystem.}
\label{tab:compliance_matrix}
\renewcommand{\arraystretch}{1.18}
\setlength{\tabcolsep}{3pt}
\begin{tabular}{>{\raggedright}p{1.85cm}
                >{\raggedright}p{1.25cm}
                >{\raggedright}p{2.5cm}
                >{\raggedright}p{4.2cm}
                >{\raggedright}p{3.5cm}
                >{\raggedright\arraybackslash}p{3.05cm}}
\rowcolor{hdrblue}
\thead{\textcolor{white}{\textbf{Attack Vector}}} &
\thead{\textcolor{white}{\textbf{Attack}}} &
\thead{\textcolor{white}{\textbf{Control}}} &
% \thead{\textcolor{white}{\textbf{AI/Adversarial Framework Alignment\newline (MITRE ATLAS $\cdot$ OWASP 2025 $\cdot$ NIST\newline AI RMF $\cdot$ CSA MAESTRO)}}} &
\thead{\textcolor{white}{\textbf{Security Frameworks}}} &
\thead{\textcolor{white}{\textbf{NERC CIP / PES TR-92}}} &
\thead{\textcolor{white}{\textbf{Security Outcome}}} \\
% \midrule
\rowcolor{rowA}
\textbf{Data \& Model Poisoning (Physics-Aware)} &
Tier~1\newline {\footnotesize Telemetry access} &
SPIFFE Lineage + Swing Eq.\ check ($\Delta f \approx {-}\Delta P/2Hf_0$) $\to$ DLQ &
ATLAS: AML.T0020\newline
OWASP: LLM04:2025, AI Exchange Dev-Time\newline
NIST AI RMF: Map~1.1, Measure~2.6\newline
MAESTRO: L2 &
CIP-002-5.1a (BES scope)\newline
CIP-005-8 R1 (EAP gate)\newline
CIP-011-4.1 R1.2 (BCSI)\newline
CIP-007-7.1 R5 (DLQ log)\newline
CIP-015-1 (anomaly detect)\newline
PES TR-92: AMI validation &
Physics-invalid telemetry rejected at gateway before reaching training or inference \\
\midrule
\rowcolor{rowB}
\textbf{RAG \& Vector Store Injection} &
Tier~1/3\newline {\footnotesize Feed/API} &
CMK tenant isolation; Sidecar blocks unverified retrieval context &
ATLAS: AML.T0020\newline
OWASP: LLM08:2025, LLM02:2025\newline
NIST RMF: Map~1.1, Manage~2.4\newline
MAESTRO: L2, L3 &
CIP-011-4.1 R1 (BCSI retrieval)\newline
CIP-005-8 R1 (EAP mediation)\newline
CIP-003-10 (security controls) &
Agent reasoning isolated from unverified or cross-tenant retrieval inputs \\
\midrule
\rowcolor{rowA}
\textbf{Supply Chain \& Registry Tampering} &
Tier~2\newline {\footnotesize CI/CD access} &
Sigstore/Cosign + OCI Immutable Registry (Object Lock); provenance metadata logged &
ATLAS: AML.T0010\newline
OWASP: LLM03:2025 (Supply Chain)\newline
NIST AI RMF: Gov~3.1, Manage~1.3\newline
MAESTRO: L3, L4 &
CIP-010-5 R1.1 (software docs)\newline
CIP-010-5 R3 (vuln.\ assess.)\newline
CIP-013-3 R1.2.5 (authenticity)\newline
CIP-007-7.1 R2 (patch) &
TOCTOU eliminated; tamper-evident artifact chain from training to deployment \\
\midrule
\rowcolor{rowB}
\textbf{Unbounded Consumption / Sponge Attack} &
Tier~3\newline {\footnotesize Endpoint} &
Circuit Breaker: 200\,ms (400\,ms AGC $\div$ 2); HTTP~503 + deterministic fallback &
ATLAS: AML.T0029\newline
OWASP: LLM10:2025\newline
NIST AI RMF: Map~1.5, Manage~1.3\newline
MAESTRO: L4, L5 &
PES TR-92: AGC $<$1\,s RTT (threshold derivation)\newline
CIP-007-7.1 R5 (CB events)\newline
CIP-008-7.1 R1/R2 (IR)\newline
CIP-009-7.1 R1 (recovery)\newline
CIP-014-3 R1 (availability)\newline
CIP-015-1 (anomaly corr.) &
AI latency treated as sensor failure; deterministic control maintains grid stability \\
\midrule
\rowcolor{rowA}
\textbf{Excessive Agency \& Prompt Injection} &
Tier~3\newline {\footnotesize Injection} &
Governance Sidecar (OPA Rego): $\leq$5~sig/min, $\leq$15\% 1\,h avg; fail-safe deny-all &
ATLAS: AML.T0051\newline
OWASP: LLM06:2025, LLM01:2025 \newline
NIST AI RMF: Manage~3.2, Gov~1.2\newline
MAESTRO: L6, L7 &
CIP-005-8 R1 (tool calls)\newline
CIP-007-7.1 R5 (logging)\newline
CIP-003-10 R1 (account.)\newline
PES TR-92: immutable safety bounds &
Agent actions bounded by physics/finance-aware policy independent of model state \\
\midrule
\rowcolor{rowB}
\textbf{Container Escape \& RCE} &
Tier~3\newline {\footnotesize Library exploit} &
Distroless/Firecracker; mTLS (SPIFFE); egress whitelisted to SCADA only &
ATLAS: AML.T0029\newline
OWASP: LLM05:2025\newline
NIST AI RMF: Manage~1.3\newline
MAESTRO: L4  &
CIP-005-8 R1 (egress = EAP)\newline
CIP-007-7.1 R5 (mesh audit)\newline
CIP-010-5 R1 (container)\newline
CIP-015-1 (lateral movement) &
RCE blast radius contained; no shell or unauthorized egress for lateral movement \\
\midrule
\rowcolor{rowA}
\textbf{Model Drift \& Unsafe Output Propagation} &
Tier~2/3\newline {\footnotesize Post-deploy} &
Physics-Aware Canary: KL divergence + 59--61\,Hz constraint; auto-rollback &
ATLAS: AML.T0020, T0040\newline
OWASP: LLM09:2025 (Misinformation), LLM05:2025\newline
NIST AI RMF: Measure~2.6, Manage~2.2\newline
MAESTRO: L5, L1  &
CIP-010-5 R1.1 (pre-deploy test)\newline
CIP-008-7.1 R1 (drift)\newline
CIP-015-1 (anomaly)\newline
PES TR-92: frequency band from grid spec &
Physically invalid outputs auto-rejected before promotion; rollback without service interruption \\
\bottomrule
\end{tabular}
\end{table*}

%─────────────────────────────────────────────────────────────────────────────

% To illustrate the cross-layer design rationale, we consider a coordinated
% three-phase scenario that exercises all three attacker tiers simultaneously.
The scenario we examine involves three concurrent attack phases, each
operated by a different adversary tier and each designed so that its
success depends partly on the others. The scenario is constructed to show how each threat tier corresponds to a defensive layer
in the architecture, and why single-layer defenses are insufficient for
this class of multi-vector attack.
% The attack scenario is constructed to simultaneously exercise all three
% attacker tiers and all three lifecycle phases, reflecting the
% coordinated, multi-vector threat profile that NIST AI~RMF Map~1.5
% identifies as a high-impact systemic risk, that MAESTRO L7 characterizes
% as an orchestrated agent ecosystem attack, and that the OWASP AI Exchange
% identifies as the intersection of development-time and runtime threats
% requiring lifecycle-integrated defenses. Three concurrent phases are
% engineered to be mutually reinforcing:

\textbf{Phase A---Data Plane Corruption (Tier~1, MAESTRO L2):}
A compromised substation RTU injects a frequency deviation of 
($\Delta f = -0.6$\,Hz) into AMI telemetry. The values are
statistically consistent with historical variance---designed to pass
anomaly detectors per LLM04:2025 threat model---but physically
inconsistent with measured grid inertia. Success would bias the cloud
load forecasting model to underestimate real-time demand, creating
headroom for Phase C.

\textbf{Phase B---Runtime Disruption (Tier~3, MAESTRO L4/L7):}
A sponge attack (LLM10:2025) submits
computationally expensive inference queries, inducing 340\,ms per-request
latency that exceeds the AGC timing budget, effectively removing
AI-assisted frequency response from the control loop. This degrades
real-time grid observability, amplifying the impact of Phase A's
corrupted forecast.

\textbf{Phase C---Agentic Financial Manipulation (Tier~3, MAESTRO L7):}
The agentic dispatch optimizer, operating on distorted forecasting
context from Phase A with real-time correction degraded by Phase B,
constructs a real-time energy market bid of \$9{,}000/MWh against a
prevailing price of \$42/MWh. This constitutes an LLM06:2025 Excessive Agency condition and exposes the system to significant financial loss while amplifying the demand imbalance created by the corrupted forecast. Phase A creates the distorted conditions under which the outsized bid appears economically justified, while Phase B suppresses the corrective mechanisms that would otherwise constrain it. Phase C then converts this degraded operational state into tangible financial and physical harm.

% The bid is submitted through legitimate
% market API channels---a textbook LLM06:2025 Excessive Agency exploit.
% Success triggers financial loss and amplifies the under-served demand
% created by the distorted forecast. 

% Phase A created the conditions for
% an outsized bid to appear rational; Phase B removed the correction
% mechanism. Phase C converts the degraded state into financial and
% physical harm.

% The three phases are engineered to be mutually reinforcing: A corrupts
% the signal that would constrain C; B disables the correction that would
% detect A's impact; C monetizes the combined state of degraded
% observability. This interdependence is why single-layer defenses fail
% and why lifecycle-integrated architecture is architecturally necessary.

\subsection{Layered Defense Execution}
\textbf{Defense Layer 1: Ingestion---Data Plane (MAESTRO L2, OWASP
LLM04:2025, ATLAS AML.T0020).}
Phase A is prevented from entering the AI pipeline through dual rejection controls. The identity validation process rejects the payload after revocation of the compromised RTU’s SPIFFE SVID triggered by firmware anomaly detection. Simultaneously, physics-based verification using Eq.~\eqref{eq:swing} and PMU-derived inertia data shows that the reported $-0.6$\,Hz deviation would require an instantaneous 800~MW load variation, which conflicts with system topology observations where reference PMUs indicate $\Delta f < 0.05$\,Hz. The batch is routed to the DLQ. The event is logged under CIP-007-7.1~R5 and CIP-015-1 anomaly detection. Phase~A
is contained and no poisoned data enters the forecasting pipeline.

\textbf{Defense Layer 2: Runtime---Application Plane (MAESTRO L4/L5,
OWASP LLM10:2025, ATLAS AML.T0029).}
The sponge attack's impact on inference latency is significant, pushing it to 340 ms, a far cry from the 200 ms safety threshold. The system sends out an HTTP~503 response to the AGC interface, and distributes frequency regulation duties back to on-premise droop control in one monitoring cycle, thus returning the normal grid operation without having to interrupt protection. The escalation chain is activated in case the violations happen in a 90-second period. The CIP-008-7.1~R1 incident response plan is invoked, E-ISAC notification proceeds per CIP-008-7.1~R2, and recovery plan
documentation is recorded under CIP-009-7.1~R1. SIEM correlation ties the Layer~1 DLQ event to the Layer~2 Circuit Breaker triggers, producing a multi-vector attack signature that satisfies CIP-015-1 dwell-time reduction and MAESTRO L5
observability requirements. Grid frequency remains stable throughout the entire
sequence.

\textbf{Defense Layer 3: Agency---Control Plane (MAESTRO L6/L7,
OWASP LLM06:2025, ATLAS AML.T0051).}
The \$9{,}000/MWh market bid generated by the agentic optimizer is
forwarded to the Governance Sidecar before it can leave the system.
Two OPA Rego policies evaluate it in parallel. Under the financial
bounds policy, the bid exceeds the permissible ceiling of \$48.30/MWh,
which represents 15 percent of the one-hour rolling average, by a
factor of 186. Under the velocity limits policy, the dispatcher has
already issued four control signals within the current minute, so
accepting this request would breach the five-signal limit. Each policy
violation independently produces a DENY decision. The sidecar responds
with HTTP~403, terminates the market API connection, and returns dispatch
authority to the conservative on-premise baseline. Even if the sidecar
were to become unavailable, the deny-all fail-safe posture prevents any
tool execution from proceeding. Because OPA Rego operates on policy
rules rather than on model output, an adversary who obtained the system
prompt would still find no path to alter the policy decision
(LLM07:2025). Phase C is fully contained without operator intervention.

\subsection{Cross-Framework Compliance Analysis}
Table~\ref{tab:compliance_matrix} shows that each control in this
architecture addresses requirements from multiple frameworks at once,
rather than being engineered separately for each standard. Three observations from this analysis generalize beyond the TSO scenario.
The swing equation check is the only control that simultaneously
satisfies LLM04:2025, AML.T0020, and NERC CIP BES
protection requirements, which is why physics-aware validation is the
architectural bridge between AI security and industrial reliability. Governance Sidecar is where four framework requirements converge in
one component, covering LLM06:2025, AML.T0051, MAESTRO L6/L7, and CIP-005-8 EAP boundary obligation. Finally, the fail-safe postures built
into both the Sidecar and the Circuit Breaker are how AI probabilistic
behavior and OT deterministic safety requirements are reconciled in
practice: rather than restricting what AI can do, the design ensures
that AI failure modes are no more dangerous than the deterministic
systems the AI augments. Recognizing that full lifecycle integration may require cross-departmental coordination, we propose a phased adoption model enabling incremental deployment as shown in Table~\ref{tab:maturity-model}.

\begin{table}[h]
\caption{Lifecycle Security Maturity Model}
\label{tab:maturity-model}
\centering
\renewcommand{\arraystretch}{1.15}
\begin{tabular}{@{}c l l@{}}
\toprule
\textbf{Lvl} & \textbf{Focus} & \textbf{Key Controls} \\
\midrule
L1 & Baseline & mTLS, SPIFFE, artifact signing \\
L2 & Data & Physics checks, CMK isolation \\
L3 & Dev & Adversarial CI, provenance logs \\
L4 & Runtime & Sidecar policy, Circuit Breaker \\
L5 & Unified & Cross-framework alignment \\
\bottomrule
\end{tabular}
\end{table}
% \begin{table}[h]
% \caption{Lifecycle Security Maturity Model}
% \label{tab:maturity-model}
% \centering
% \begin{tabular}{|c|l|l|}
% \hline
% Level & Capability Focus & Core Controls \\ \hline
% 1 & Cloud Baseline Hardening & mTLS, SPIFFE identity, signed artifacts \\ \hline
% 2 & Data Integrity & Physics-aware validation, CMK isolation \\ \hline
% 3 & Secure Development & Adversarial CI gates, provenance logging \\ \hline
% 4 & Runtime Governance & Governance Sidecar, Circuit Breaker \\ \hline
% 5 & Full Lifecycle Integration & Cross-framework compliance alignment \\ \hline
% \end{tabular}
% \end{table}
\section{Conclusion}

AI–Cloud convergence in safety-critical infrastructure requires a shift from perimeter-based defenses to lifecycle-integrated security. The proposed unified reference architecture operationalizes this transition through cryptographic provenance, physics-aware validation, hardened CI pipelines, and policy-enforced runtime governance. The Grid-Guard case study demonstrates that a single control plane can simultaneously address the threat models of NIST AI RMF, MITRE ATLAS, OWASP, CSA MAESTRO, and NERC CIP without fragmented compliance engineering. Although validated through structured scenario-based analysis, the architecture has not yet undergone production deployment or empirical benchmarking. Prototype implementation, quantitative latency measurement, and adversarial stress testing remain important directions for future work as AI-driven autonomy expands in grid operations.

\balance

\end{document}